\documentclass[pra,twocolumn,superscriptaddress,nofootinbib,showpacs]{
revtex4}
\usepackage{bm}
\usepackage{graphicx}
\newcommand{\rrangle}{\rangle\!\rangle}
\newcommand{\0}{|0\rangle\!\rangle}
\newcommand{\1}{|1\rangle\!\rangle}
\newcommand{\s}{|s\rangle\!\rangle}
\newcommand{\x}{|x\rangle\!\rangle}

\newcommand{\arr}{|a\rangle\!\rangle}
\newcommand{\brr}{|b\rangle\!\rangle}

\newcommand{\A}{{\mathcal A}}
\newcommand{\B}{{\mathcal B}}

\newcommand{\K}{{\mathcal K}}

\begin{document}
\title{\bf Quantum vs classical computation: a proposal opening a new
perspective}
\author{Angelo Bassi}
\email{bassi@ictp.trieste.it} \affiliation{The Abdus Salam
International Centre for Theoretical Physics, Trieste, and \\
Istituto Nazionale di Fisica Nucleare, sezione di Trieste, Italy.}
\author{GianCarlo Ghirardi}
\email{ghirardi@ts.infn.it} \affiliation{Department of Theoretical
Physics, University of Trieste, \\ The Abdus Salam International
Centre for Theoretical Physics, Trieste, and \\ Istituto Nazionale
di Fisica Nucleare, sezione di Trieste, Italy.}
\begin{abstract}
We develop a classical model of computation (the S model) which
captures some important features of quantum computation, and which
allows to design fast algorithms for solving specific problems. In
particular, we show that Deutsch's problem can be trated within
the S model of computation in the same way as within quantum
computation; also Grover's search problem of an unsorted database
finds a surprisingly fast solution. The correct understanding of
these results put into a new perspective the relationship between
quantum and classical computation.
\end{abstract}
 \maketitle

\section{Introduction}

The intensive research of the past decade has shown that quantum
computation \cite{ben,fey,deu,divt,bent,nc} can solve some
problems more efficiently than classical computation: the two most
striking example are Shor's algorithm \cite{sho,ejo,exp3} for
factoring large integers, which is exponentially faster than any
classical algorithm so far developed, and Grover's search
algorithm \cite{gro,bb,bv}, which is quadratically faster than the
corresponding classical one. The underlying reason for the power
of quantum computers is still under debate, but the general
feeling is that entanglement plays a fundamental role
\cite{joz,ej,lp}.

The procedure for translating a classical computational problem
into a quantum mechanical one is well established and it goes as
follows. Consider a problem of classical computation expressed by
a function from $n$ bits to $m$ bits\footnote{$\B_{i}$ denotes the
set of all $i$--bit--long strings.}:
\begin{equation} \label{cnrf}
f(x): \; \B_{n} \; \longrightarrow \; \B_{m};
\end{equation}
since quantum algorithms are necessarily given in terms of unitary
--- thus reversible --- operations, one has first of all to
rephrase the above problem in a reversible way. There is a
standard procedure for doing so, which does not alter the
complexity of the problem \cite{lec,bene} and consists in
replacing the function $f(x)$ with the new function:
\begin{eqnarray} \label{crf}
F(x,y): {\mathcal B}_{n}\times {\mathcal B}_{m} & \longrightarrow
& {\mathcal B}_{n} \times {\mathcal B}_{m}
\nonumber \\
(x,y) & \longrightarrow  & (x,y\oplus f(x))
\end{eqnarray}
($\times$ denotes the cartesian product between two sets and
$\oplus$ the bit--wise addition modulo 2). Note that the problem
is the same as the original one, as the first $n$ bits added to
the output are equal to the corresponding input bits and thus give
no extra information on the properties of the function $f(x)$,
whose precise form is determined only by the remaining last $m$
bits; in particular, if $y$ is initially set to $00\ldots 0$ ($m$
times), then the output gives directly the value $f(x)$.

The quantum mechanical translation of the function $F(x,y)$ is now
straightforward. Consider the unitary operator $U$, defined on the
computational basis of the tensor product Hilbert
space\footnote{The Hilbert space ${\mathcal H}_{i}$ has dimension
equal to $2^i$.} ${\mathcal H}_{n}\otimes {\mathcal H}_{m}$, in
terms of the function $F(x,y)$ as follows:
\begin{eqnarray} \label{uop}
U: {\mathcal H}_{n}\otimes {\mathcal H}_{m} & \longrightarrow
& {\mathcal H}_{n} \otimes {\mathcal H}_{m} \nonumber \\
|x\rangle \otimes |y\rangle & \longrightarrow & |x\rangle \otimes
|y \oplus f(x)\rangle.
\end{eqnarray}
Given the quantum circuit implementing the operator $U$, one can
search for quick algorithms for the solution of the problem.

The reason behind the success of algorithms such as those of Shor
and Grover is the following: while, classically, we can input only
one value at a time into the circuit implementing the function
$f(x)$, quantum mechanically we can do much better; by preparing
the input state of the first $n$ qubits in a superposition of all
computational basis states ($N = 2^n$):
\begin{equation} \label{aqq}
|\psi\rangle \; = \; \frac{1}{\sqrt{N}}\;\sum_{x = 0}^{N-1} \;
|x\rangle,
\end{equation}
and setting (for simplicity) $y = 0$, then, in virtue of the
linear character of quantum operators, we get as the output of the
quantum circuit:
\begin{equation} \label{sup}
U\,|\psi\rangle\otimes|0\rangle \; = \;
\frac{1}{\sqrt{N}}\;\sum_{x = 0}^{N-1} \; |x\rangle \otimes
|f(x)\rangle.
\end{equation}
With a single step, we have been able to compute {\it all} values
of the function $f(x)$. Of course, we do not have direct access to
any precise state of a superposition like (\ref{sup});
nevertheless, with appropriate manipulations, it is possible to
extract the desired information in a rather quick way: this,
basically, is the goal of quantum algorithm design.

In this paper, we briefly review how a computational problem is
treated both within classical and quantum computation. We then
formulate a new classical model of computation (the S model) which
captures some features of quantum computation, in particular the
possibility of inputting superposition (to be understood in an
appropriate way) of states: thanks to this property, we will show
that some classical problems can be solved in a surprisingly fast
way. The correct understanding of this result yields a new
perspective on the relationship between quantum and classical
computation.

\section{Computational problems}

Within computational complexity theory \cite{pap}, problems are
formalized by resorting to the formal--language theory; for the
purposes of this article, a much simpler approach is sufficient:
we define a computational problem as that of {\it finding a
specific property of a given function}
\begin{equation} \label{eq7}
f(x) : \B_{n} \; \longrightarrow \; \B_{m},
\end{equation}
taking $n$ bits into $m$ bits. A typical example is the SAT
problem: given any Boolean function (i.e. a function whose
variables are connected only by $\wedge$ (AND), $\vee$ (OR),
$\neg$ (NOT) Boolean connectives) taking $n$ bits into 1 bit,
e.g.:
\begin{equation}
f(x_{1},x_{2},x_{3}) \; = \; (\neg x_{1} \vee x_{2} \vee \neg
x_{2}) \wedge (x_{1} \vee \neg x_{2} \vee x_{3})
\end{equation}
we have to find whether there is an input value whose output is 1.

\subsection{Solving problems within classical computation}

It is well known that any function of the type (\ref{eq7}) is {\it
computable}, i.e. there exists a circuit such that, given $x$ as
input, it outputs the value $f(x)$. It is convenient to divide the
procedure for finding a solution of a computational problem into
the following two steps:

\noindent 1) Given a function $f(x)$ of the type (\ref{eq7}), one
first constructs the circuit implementing it.

\noindent 2) Given the circuit, one works out an algorithm for
finding the solution of the problem.

Needless to say, the complexity of the global procedure for
solving the problem must take into account both the number of
steps required to construct the circuit implementing the function
$f(x)$ and the complexity of the algorithm which, by resorting to
such a circuit, solves the problem. Anyway, for the most important
classical computational problems, the implementation of the
circuit poses no problems (i.e. the procedure required for its
construction is {\it polynomial} in the size --- defined in an
appropriate way --- of the problem). For this reason, one can
usually focus his attention only on the complexity of algorithms.

For example, with reference to the SAT problem, the construction
of the circuit is {\it polynomial} in the size of the problem
(i.e. the number of Boolean connectives); on the contrary, no
algorithm for solving the problem is known, which is polynomial.

\subsection{Solving problems within quantum computation}

In analogy with the classical situation, the procedure for solving
a computational problem within quantum computation can be divided
into two steps.

\noindent 1) Given a computable function of the type (\ref{eq7}),
one first constructs the quantum circuit implementing the unitary
operator
\begin{eqnarray} \label{cqa}
U: \, {\mathcal H}_{n} \otimes {\mathcal H}_{m} & \longrightarrow
& {\mathcal H}_{n} \otimes {\mathcal H}_{m} \nonumber \\
|x\rangle \otimes |y\rangle & \longrightarrow & |x\rangle \otimes
|y \oplus f(x)\rangle.
\end{eqnarray}

\noindent 2) Given the quantum circuit, one works out appropriate
algorithms for finding the solution of the problem.

A remarkable property of quantum computation is the following: it
has been proved (see, e.g. \cite{nc}) that there is a general
procedure for translating any classical circuit into the
corresponding quantum circuit (in the sense of (\ref{cqa})), which
is {\it polynomial} in the size of the circuit, i.e. in the number
of elementary gates appearing in it. Accordingly, the complexity
of the construction of a circuit is the same, within classical and
within quantum computation. What marks the difference between the
two theories is the possibility to work out quantum algorithms for
solving specific problems which are faster than any known
classical algorithm that solves the same problem.

\section{The S model of computation}

In this section we define a new model of computation, which we
call the S model: its building blocks are just three states, $\0,
\1$ and $\s$. The state $\s$ will formally play the role of the
{\it superposition}, to be understood in an appropriate way, of
states $\0$ and $\1$. In the subsequent sections we will analyze
how the S model can be used to solve certain computational
problems.

\subsection{The sbit}

\noindent {\it Single sbit.} Consider a set $\K_ 1$ containing
three elements $\0$, $\1$, $\s$: such elements represent the
possible states of a single sbit. In $\K_ 1$ we define a {\it sum}
$+$ according to the following rules:
\begin{center}
\begin{tabular}{c|ccc}
$+$  & $\0$ & $\1$ & $\s$ \\ \hline
$\0$ & $\0$ & $\s$ & $\s$ \\
$\1$ & $\s$ & $\1$ & $\s$ \\
$\s$ & $\s$ & $\s$ & $\s$
\end{tabular}
\end{center}
It is easy to see that this sum is {\it commutative} and {\it
associative}, and these are the only two properties which we are
interested in: thanks to them we can write any sum of three or
more elements without specifying the order in which the sum is
performed.

We define the {\it computational basis} as the {\it smallest}
subset of $\K_ 1$ such that every element of $\K_ 1$ can be
written as a sum of elements of the computational basis: in the
present case, the computational basis is simply $\{ \0, \1 \}$ and
has the same cardinality as $\B_{1}$.

\noindent {\it Multiple sbits.} The state space $\K_ {n}$ of $n$
sbits is, like in classical computation, the cartesian product of
$n$ sets $\K_ {1}$, i.e. the set of all strings of the form:
\begin{equation}
\underbrace{\0 \1 \0 \s \ldots \s}_{n \; \makebox{\small
elements}} \quad \equiv \quad |010 \makebox{s} \ldots \makebox{s}
\rrangle.
\end{equation}
$\K_ {n}$ contains $3^n$ elements and, in it, we define a sum $+$
as the sbit--wise sum of the two elements being added, like e.g.
\begin{eqnarray}
|01\makebox{s}\rrangle + |110\rrangle & = & (\0 + \1)(\1 + \1)(\s
+ \0) \nonumber \\
& = & |\makebox{s}1\makebox{s}\rrangle.
\end{eqnarray}
The sum identifies a computational basis in $\K_ {n}$ defined, as
before, as the smallest subset\footnote{One can prove that such a
set exists and is unique.} such that every element of $\K_ {n}$
can be written as a sum of elements of the computational basis; it
is not difficult to check that the computational basis states are
only those which do not contain $\s$ sbits: there are $2^n$ states
of this kind, which can be trivially set into a one--to--one
correspondence with the bits of ${\mathcal B}_{n}$.

\subsection{Operations with sbits}

In analogy with the quantum situation, one would be tempted to
consider as ``valid'' operations only additive functions on $\K_
{n}$, i.e. those which, given two elements $|a\rrangle$ and
$|b\rrangle$ and their images, act as follows:
\begin{equation} \label{lin}
G\,[\arr  + \brr ] \; = \; G\,\arr  + G\,\brr;
\end{equation}
however this is not possible, as request (\ref{lin}) leads to
inconsistencies. As an example, consider an operator $G: \K_ {2}
\rightarrow \K_ {1}$ which acts on the computational basis as
follows:
\begin{eqnarray} \label{example}
G\, |00\rrangle & = & \0, \quad G\, |01\rrangle = \1, \nonumber \\
G\, |10\rrangle & = & \0, \quad G\, |11\rrangle = \0.
\end{eqnarray}
Since $|ss\rrangle = |00\rrangle + |01\rrangle + |10\rrangle +
|11\rrangle = |00\rrangle + |10\rrangle + |11\rrangle =
|00\rrangle + |11\rrangle$, if $G$ satisfied equation (\ref{lin}),
we would have, at the same time, both $G\, |ss\rrangle = \s$ and
$G\, |ss\rrangle = \0$, which is not consistent.

To avoid problems of this kind, we adopt the following strategy.
Consider a generic sbit $\arr  \in \K_ {n}$, and the set $\A$
defined as the {\it biggest} subset (of course, in counting the
elements of the subset, one takes into account only different
elements) of ${\mathcal B}_{n}$ such that\footnote{In the
following, it is understood that when we write a sum like
(\ref{defa}), the set $\A$ appearing in it is always the maximal
set in the sense given above.}:
\begin{equation} \label{defa}
\arr  \; = \; \sum_{x \in \A} \x;
\end{equation}
such a set exists, because of the way in which the computational
basis is defined, and it is unique. It is easy to check that,
given a multiple sbit $\arr$, e.g. $\arr = |010s \ldots s10 \ldots
\rrangle$, for which $s$ appears $k$ times, the maximal set $\A$
associated to it contains precisely $2^k$ computational basis
sbits, which are obtained by replacing in all possible ways the
$s$' of the ordered sequence $010s \ldots s10 \ldots$ defining
$\arr$ by 0's and 1's, while the 0's and 1's appearing in original
sequence are kept fixed.

We can now define the operations on sbits which are allowed as
those that satisfy the requirement
\begin{equation}\label{sum}
G\,\arr  \; = \; \sum_{x \in \A} G \, \x,
\end{equation}
for any state $\arr $ belonging to $\K_ n$, where $\A$ is the
maximal set associated to $\arr$. We call this condition {\it
weak--additivity}, and the operations satisfying it will be called
weakly--additive (w--additive)\footnote{Obviously, w--additivity
does not imply additivity; as an example, consider once more the
operator defined in (\ref{example}), and assume it to be
w--additive: then, we have that $|ss\rrangle = |00\rrangle +
|11\rrangle$, but $G\, |ss\rangle = \s \neq G\, |00\rrangle +
G\,|11\rrangle = \0$.}.

Note that the number of terms appearing in the sum (\ref{sum})
defining the w--additivity property --- just as the number of
terms appearing in equation (\ref{aqq}) --- in general grows {\it
exponentially} with the number of input sbits. Anyway, as it will
be clear from the examples of the following sections, there are
w--additive circuit implementing w--additive operators, whose size
(number of elementary gates) is {\it polynomial} in the number of
input sbits, yet they compute sums of the type (\ref{sum}),
without involving loops or any kind of hidden exponential slowdown
in the computational time.

We now introduce some elementary w--additive gates, starting with
the simplest ones, which take one sbit into one sbit. There are
only two interesting gates of this kind, the NOT and H gates;
their action on the computational basis is:
\begin{center}
NOT =
\begin{tabular}{c|c}
input & output \\\hline
$\0$ & $\1$ \\
$\1$ & $\0$
\end{tabular}
\hspace{.5cm} H =
\begin{tabular}{c|c}
input & output \\\hline
$\;\0$ & $\s$ \\
$\;\1$ & $\s$
\end{tabular}
\end{center}
The NOT gates simply acts as its classical counterpart, while the
H gate in some sense mimics the quantum Hadamard gate, since it
takes both computational basis states in $\s$, which can be seen
as the sum of $\0$ and $\1$.

In a similar manner, we can introduce gates such as the
w--additive AND, OR, FANOUT, which are defined on the
computational basis of the appropriate domains as their classical
counterparts\footnote{We stress that, due to w--additivity, we
need to define the gates only on the computational basis, and they
turn out to be automatically defined on the whole domain.}. In
appendix 1, it is proved that there exists a {\it universal set}
of w--additive gates, i.e. a fixed number of elementary
w--additive gates which can be used to compute any other
w--additive operation. Appendix 2 shows other important elementary
gates; we call in particular the attention of the reader on the
two w--additive gates $C_{0}$ and $C_{1}$, for the role they will
play in what follows: $C_{0}$ outputs the sbit $|0\rrangle$ for
any input value, while $C_{1}$ outputs always the sbit
$|1\rrangle$.

It is important to stress that, while within quantum computation
the product of two unitary operators is always a unitary operator,
it is not always true that the successive application of two
w--additive operations still gives a w--additive operation;
accordingly, when we combine w--additive gates to implement a
function, we have to check that the so obtained circuit is
w--additive. This represents the most serious limitation of the S
model of computation but, as it will become clear in what follows,
this limitation does not make the model useless. Actually, in
various interesting cases it is possible to show how to combine
elementary w--additive gates to automatically obtain w--additive
circuits which can be used to solve interesting instances of
important computational problems. As a matter of fact, it is
possible to prove some theorems stating sufficient conditions for
a circuit to be w--additive: here we propose two of them (the
proof of the first theorem is given in the appendix; the proof of
the second one is left to the reader).

\noindent {\it Theorem 1.} Let $G_{1} : \K_ {n} \rightarrow \K_
{1}$ and $G_{2} : \K_ {m} \rightarrow \K_ {1}$ be two w--additive
operations. Then $G : \K_ {n} \times \K_ {m} \rightarrow \K_ {1}$
defined as $G \equiv$ AND $[ G_{1} \times G_{2} ]$ is also
w--additive.
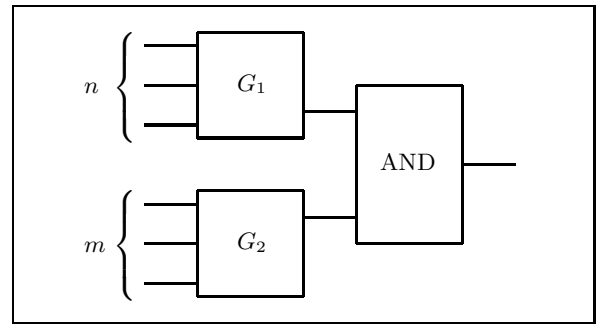
\begin{figure}
\begin{center}
\begin{picture}(220,120)(0,0)
\put(0,0){\line(1,0){220}} \put(0,120){\line(1,0){220}}
\put(0,0){\line(0,1){120}} \put(220,0){\line(0,1){120}}
\thicklines \put(70,10){\line(1,0){40}}
\put(70,50){\line(1,0){40}} \put(70,10){\line(0,1){40}}
\put(110,10){\line(0,1){40}}
\put(70,70){\line(1,0){40}} \put(70,110){\line(1,0){40}}
\put(70,70){\line(0,1){40}} \put(110,70){\line(0,1){40}}
\put(50,15){\line(1,0){20}} \put(50,30){\line(1,0){20}}
\put(50,45){\line(1,0){20}} \put(50,75){\line(1,0){20}}
\put(50,90){\line(1,0){20}} \put(50,105){\line(1,0){20}}
\put(110,40){\line(1,0){20}} \put(110,80){\line(1,0){20}}
\put(130,30){\line(1,0){40}} \put(130,90){\line(1,0){40}}
\put(130,30){\line(0,1){60}} \put(170,30){\line(0,1){60}}
\put(170,60){\line(1,0){20}}
\put(38,87){$\left\{ \makebox(8,25)[t]{$ $} \right.$}
\put(38,27){$\left\{ \makebox(8,25)[t]{$ $} \right.$}
\put(27,87){$n$} \put(27,27){$m$} \put(85,88){$G_{1}$}
\put(85,28){$G_{2}$} \put(139,58){AND}
\end{picture}
\caption{Theorem 1: if $G_{1}$ and $G_{2}$ are w--additive, then
the global circuit depicted in the figure is w--additive.}
\end{center}
\end{figure}

\noindent {\it Theorem 2.} Let $G_{1} : \K_ {n} \rightarrow \K_
{1}$ be a w--additive operation. Then $G : \K_ {n} \rightarrow \K_
{1}$ defined as $G \equiv$ NOT $[G_{1}]$ is also w--additive.
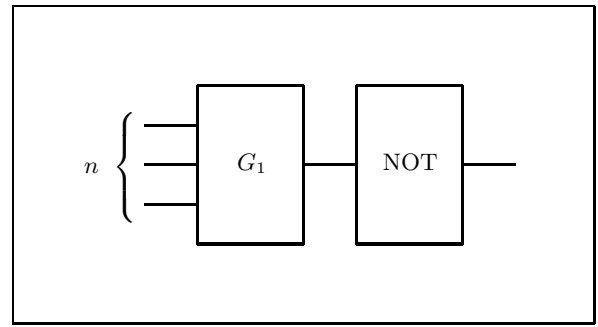
\begin{figure}
\begin{center}
\begin{picture}(220,120)(0,0)
\put(0,0){\line(1,0){220}} \put(0,120){\line(1,0){220}}
\put(0,0){\line(0,1){120}} \put(220,0){\line(0,1){120}}
\thicklines \put(70,30){\line(1,0){40}}
\put(70,90){\line(1,0){40}} \put(70,30){\line(0,1){60}}
\put(110,30){\line(0,1){60}} \put(170,60){\line(1,0){20}}
\put(50,45){\line(1,0){20}} \put(50,60){\line(1,0){20}}
\put(50,75){\line(1,0){20}}
\put(110,60){\line(1,0){20}} \put(130,30){\line(1,0){40}}
\put(130,90){\line(1,0){40}} \put(130,30){\line(0,1){60}}
\put(170,30){\line(0,1){60}} \put(170,60){\line(1,0){20}}
\put(38,57){$\left\{ \makebox(8,25)[t]{$ $} \right.$}
\put(27,57){$n$} \put(85,58){$G_{1}$} \put(140,58){NOT}
\end{picture}
\caption{Theorem 2: if $G_{1}$ is w--additive, then the global
circuit is also w--additive.}
\end{center}
\end{figure}

The above two theorems have an important consequence: given a
classical circuit composed of AND and NOT gates (thus also OR
gates, but not FANOUTs), the corresponding w--additive circuit
(i.e. the w--additive circuit which, on the computational basis,
acts as the classical one) can be obtained by substituting to
every classical elementary gate the corresponding w--additive
gate. This procedure, of course, is {\it linear} in the number of
elementary gates of the circuit.

When FANOUT gates are present, the situation becomes more
delicate. There are particular, but important, cases in which,
given a circuit containing one (or more) FANOUT, it is easy to
construct the corresponding w--additive circuit. Two such cases
are presented in Fig. \ref{s0} and \ref{s1}: when, in a classical
circuit, a bit is copied with a FANOUT, one of the two copies goes
through a NOT gate and, subsequently, the two copies are jointed
in an AND gate, then the w--additive gate corresponding to this
piece of the classical circuit is the $C_{0}$ gate. If, instead of
the AND gate, an OR gate is present, then the corresponding
w--additive gate is the $C_{1}$ gate.
\begin{figure}
\begin{center}
\begin{picture}(220,120)(0,0)
\put(0,0){\line(1,0){220}} \put(0,120){\line(1,0){220}}
\put(0,0){\line(0,1){120}} \put(220,0){\line(0,1){120}}
\thicklines \put(40,100){\line(1,0){100}}
\put(80,80){\line(1,0){20}} \put(100,70){\line(0,1){20}}
\put(120,80){\line(-2,1){20}} \put(120,80){\line(-2,-1){20}}
\put(120,80){\line(1,0){20}} \put(140,75){\line(0,1){30}}
\qbezier(140,75)(190,90)(140,105) \put(165,90){\line(1,0){20}}
\put(80,100){\line(0,-1){20}} \put(80,100){\circle*{5}}
\put(80,30){\line(1,0){20}} \put(100,20){\line(1,0){20}}
\put(100,40){\line(1,0){20}} \put(100,20){\line(0,1){20}}
\put(120,20){\line(0,1){20}} \put(120,30){\line(1,0){20}}
\put(106,27){$C_{0}$}
\put(10,50){Piece of a classical}  \put(10,40){circuit}
\put(150,50){Corresponding} \put(150,40){w--additive}
\put(150,30){piece}
\thinlines \qbezier(20,60)(20,80)(55,85)
\put(55,85){\vector(4,1){1}} \qbezier(170,25)(170,0)(140,20)
\put(140,20){\vector(-1,1){1}}
\end{picture}
\caption{Piece of a classical circuit where a bit is copied, one
of the two copies goes through a NOT gate, and finally the two
copies are feeded into an AND gate. The corresponding w--additive
gate is $C_{0}$.} \label{s0}
\end{center}
\end{figure}
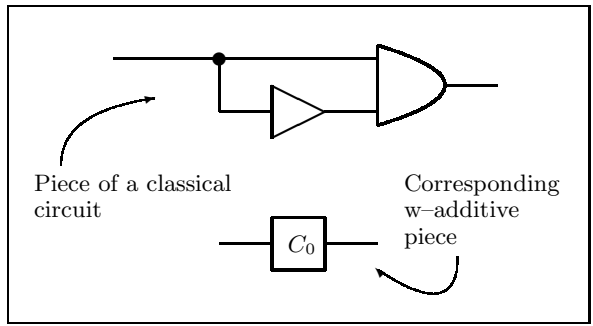
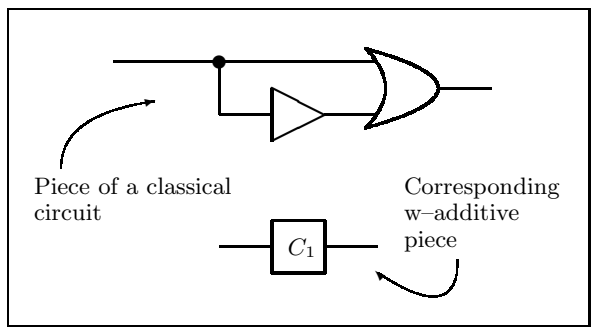
\begin{figure}
\begin{center}
\begin{picture}(220,120)(0,0)
\put(0,0){\line(1,0){220}} \put(0,120){\line(1,0){220}}
\put(0,0){\line(0,1){120}} \put(220,0){\line(0,1){120}}
\thicklines \put(40,100){\line(1,0){100}}
\put(80,80){\line(1,0){20}} \put(100,70){\line(0,1){20}}
\put(120,80){\line(-2,1){20}} \put(120,80){\line(-2,-1){20}}
\put(120,80){\line(1,0){20}} \qbezier(135,75)(150,90)(135,105)
\qbezier(135,75)(190,90)(135,105) \put(163,90){\line(1,0){20}}
\put(80,100){\line(0,-1){20}} \put(80,100){\circle*{5}}
\put(80,30){\line(1,0){20}} \put(100,20){\line(1,0){20}}
\put(100,40){\line(1,0){20}} \put(100,20){\line(0,1){20}}
\put(120,20){\line(0,1){20}} \put(120,30){\line(1,0){20}}
\put(106,27){$C_{1}$}
\put(10,50){Piece of a classical}  \put(10,40){circuit}
\put(150,50){Corresponding} \put(150,40){w--additive}
\put(150,30){piece}
\thinlines \qbezier(20,60)(20,80)(55,85)
\put(55,85){\vector(4,1){1}} \qbezier(170,25)(170,0)(140,20)
\put(140,20){\vector(-1,1){1}}
\end{picture}

\caption{Piece of a classical circuit where a bit is copied, one
of the two copies goes through a NOT gate, and finally the two
copies are feeded into an OR gate. The corresponding w--additive
gate is $C_{1}$.} \label{s1}
\end{center}
\end{figure}

We introduce now the following definition, which will play a
crucial role for the following discussion.

\noindent{\it Definition:} a classical circuit is said to be {\it
convertible} if there is an efficient (i.e. polynomial in the
number of elementary gates) procedure for converting it into the
corresponding w--additive circuit.

The previous analysis has shown that any circuit composed of AND,
OR, NOT gates and FANOUT gates appearing in a configuration like
that of Figs. \ref{s0} or \ref{s1}, is convertible. Of course, the
class of convertible circuits is much bigger: it is an open
question to ascertain how big it is.

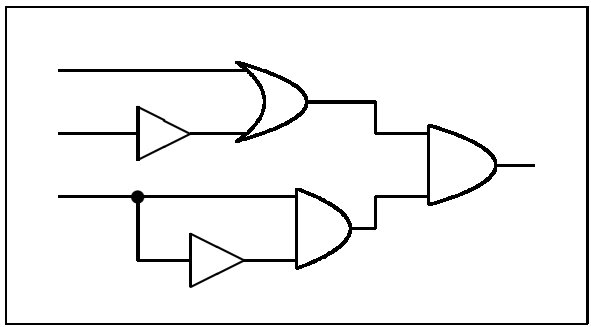
\begin{figure}
\begin{center}
\begin{picture}(220,120)(0,0)
\put(0,0){\line(1,0){220}} \put(0,120){\line(1,0){220}}
\put(0,0){\line(0,1){120}} \put(220,0){\line(0,1){120}}
\thicklines \put(20,96){\line(1,0){72}}
\put(20,72){\line(1,0){30}} \put(50,62){\line(0,1){20}}
\put(70,72){\line(-2,1){20}} \put(70,72){\line(-2,-1){20}}
\put(70,72){\line(1,0){21}} \qbezier(87,99)(108,84)(87,69)
\qbezier(87,99)(140,84)(87,69) \put(114,84){\line(1,0){26}}
\put(20,48){\line(1,0){90}} \put(50,48){\line(0,-1){24}}
\put(50,24){\line(1,0){20}} \put(70,14){\line(0,1){20}}
\put(90,24){\line(-2,1){20}} \put(90,24){\line(-2,-1){20}}
\put(50,48){\circle*{5}} \put(90,24){\line(1,0){20}}
\put(110,21){\line(0,1){30}} \qbezier(110,51)(150,36)(110,21)
\put(130,36){\line(1,0){10}}
\put(140,36){\line(0,1){12}} \put(140,84){\line(0,-1){12}}
\put(140,48){\line(1,0){20}} \put(140,72){\line(1,0){20}}
\put(160,45){\line(0,1){30}} \qbezier(160,45)(210,60)(160,75)
\put(185,60){\line(1,0){15}}
\end{picture}
\caption{Example of a convertible classical circuit.} \label{ex}
\end{center}
\end{figure}

\begin{figure}
\begin{center}
\begin{picture}(220,120)(0,0)
\put(0,0){\line(1,0){220}} \put(0,120){\line(1,0){220}}
\put(0,0){\line(0,1){120}} \put(220,0){\line(0,1){120}}
\thicklines \put(20,90){\line(1,0){100}}
\put(20,60){\line(1,0){20}} \put(40,45){\line(1,0){30}}
\put(40,75){\line(1,0){30}} \put(40,45){\line(0,1){30}}
\put(70,45){\line(0,1){30}} \put(70,60){\line(1,0){50}}
\put(45,57){NOT}
\put(20,30){\line(1,0){60}} \put(80,15){\line(1,0){30}}
\put(80,45){\line(1,0){30}} \put(80,15){\line(0,1){30}}
\put(110,15){\line(0,1){30}} \put(110,30){\line(1,0){50}}
\put(90,28){$C_{0}$}
\put(120,50){\line(1,0){30}} \put(120,100){\line(1,0){30}}
\put(120,50){\line(0,1){50}} \put(150,50){\line(0,1){50}}
\put(150,60){\line(1,0){10}} \put(128,73){OR}
\put(160,20){\line(1,0){30}} \put(160,70){\line(1,0){30}}
\put(160,20){\line(0,1){50}} \put(190,20){\line(0,1){50}}
\put(190,60){\line(1,0){15}} \put(165,42){AND}
\end{picture}
\caption{W--additive circuit corresponding to the classical
circuit of Fig. \ref{ex}.} \label{wex}
\end{center}
\end{figure}
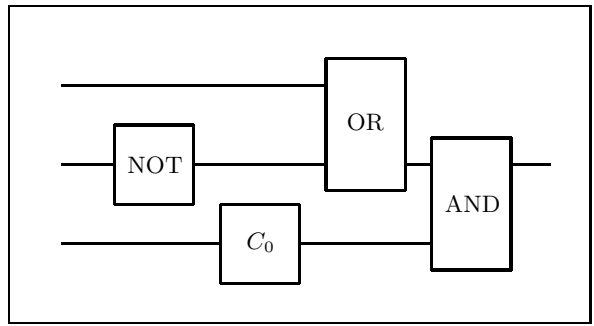

We conclude this section by giving a simple example of a
convertible circuit. Consider the classical circuit depicted in
Fig. \ref{ex}.
It is composed of AND, NOT, OR gates, and the only FANOUT appears
in a configuration like that depicted in Fig. \ref{s1};
accordingly, the circuit is convertible and the corresponding
w--additive circuit, obtained by substituting to every elementary
part of the circuit the corresponding w--additive one, is shown in
Fig. \ref{wex}.
We stress once more that:

\noindent 1) The number of steps needed to build up the
w--additive circuit is {\it proportional} to the number of
elementary gates, since the procedures require, basically, to
substitute every component of the classical circuit with the
corresponding w--additive component.

\noindent 2) The resulting w--additive circuit acts on the
computational basis of ${\mathcal K}_{3}$ as its classical
counterpart. This property follows automatically from the way it
has been constructed.

\noindent 2) Being w--additive, the new circuit admits as input
states also ``superpositions'' of the type $|ss0\rrangle$: like in
the quantum case, the output for such input states is computed
automatically by the w--additive circuit with just one single
query.

To summarize, we have defined a consistent computational model
which tries to capture some features of quantum computation: the
building block is the sbit which has two computational basis
states plus a third state which can be seen as the superposition
of the two basis states. Operations on sbits are defined on the
computational basis and the requirement of weak additivity (the
analog of linearity within quantum computation) defines in a
unique way their action for all other states; although the
combination of two w--additive operations is not always
w--additive, all convertible classical circuits can be easily
turned into the corresponding w--additive circuits. Finally, there
exists a universal set of w--additive gates.

\section{Algorithms with sbits}

We now analyze some computational problems which have particular
relevance for quantum computation. As we did in the classical and
quantum case, we divide the discussion into two parts: the
construction of the (w--additive) circuit which implements the
function defining the problem, and the formulation of the
algorithm for solving the problem. In this section we discuss this
second part, i.e. the formulation of the algorithm. We will tackle
the problem of the construction of the corresponding circuits in
the next section.

\subsection{Deutsch--Jozsa problem}

The Deutsch--Jozsa problem \cite{deu,dj,exp1,exp2} has a special
role within quantum computation, first because it was the first
problem which was proven to be easily solvable by quantum
computers, second because, being rather easy, it has become the
paradigmatic example of the way in which quantum computers work.

The problem is the following one: we are given a function $f(x):
\B_{n} \rightarrow \B_{1}$ which is either constant or
balanced\footnote{A function is balanced when it assumes the same
value for half the elements of its domain, and the other value for
the remaining ones.}; we have to decide whether $f(x)$ is constant
or balanced.

Any classical deterministic algorithm requires, in the worst case,
$2^{n-1} +1$ computations of the function $f(x)$ --- i.e. queries
to the circuit implementing it --- to check whether it is constant
or balanced; in contrast, the quantum Deutsch--Jozsa algorithm
gives the correct solution of the problem with just one query.
Without repeating the argument for the quantum case, we pass
directly to discuss the situation within the sbit model of
computation.

\noindent {\it Sbit computational problem.} In analogy with the
quantum situation, consider the w--additive circuit $G: \K_ {n}
\rightarrow \K_ {1}$, mapping the computational basis state $\x$
of $\K_ {n}$ into $|f(x)\rrangle$. The algorithm for determining
whether $f$ is constant or balanced works as follows.

{\it Step 1.} Prepare $n$ sbits, all in state $\0$.

{\it Step 2.} Apply a H gate to each sbit:
\begin{equation}
|000\ldots 0\rrangle \; \longrightarrow \; |sss \ldots s\rrangle
\; = \; \sum_{x\in \B_{n}} \x.
\end{equation}

{\it Step 3.} Make a call to the circuit:
\begin{equation} \label{dfg}
\sum_{x\in \B_{n}} \x \; \longrightarrow \; \sum_{x\in \B_{n}} G\,
\x \; = \; \left\{
\begin{array}{ll}
\0 & \makebox{if $f(x) = 0 \; \forall x$,} \\
\1 & \makebox{if $f(x) = 1 \; \forall x$,} \\
\s & \makebox{if $f$ is balanced.}
\end{array} \right.
\end{equation}

{\it Step 4.} Make a measurement: the outcome will immediately
reveal whether the function $f$ is constant or balanced and, if it
is constant, whether it is equal to $0$ or $1$.

With just a {\it single} query to the w--additive circuit, we get
the solution of the problem. Note that, once given the circuit
implementing $G$, the resources required by the algorithm are
polynomial in the size of the problem, since only $n$ sbits, $n$ H
gates and one measurement are needed. Note also that the above
algorithm can be employed to solve the more general and harder
problem of checking whether a given function is constant or not:
also in this case, a single query to the w--additive circuit is
sufficient to distinguish a constant from a non constant function
since, if $f(x)$ is non constant, in the second sum of equation
(\ref{dfg}) both terms $|0\rrangle$ and $|1\rrangle$ appear and
the output is $|s\rrangle$.

\subsection{Grover's search problem}

Consider a function which is constant everywhere in its domain,
except for one point $a$:
\begin{equation} \label{csp}
f(x): {\mathcal B}_{n} \; \longrightarrow \; {\mathcal B}_{1}
\qquad\quad f(x) \, = \, \left\{
\begin{array}{cc}
0 & \makebox{if $x \neq a$,} \\
1 & \makebox{if $x = a$:}
\end{array} \right.
\end{equation}
the problem is to identify $a$.

We know that, within classical computation, the problem cannot be
solved, on the average, with less than $N/2$ computations of the
circuit implementing the function (\ref{csp}); quantum
mechanically we can do better, as Grover proved that only $\sim
\sqrt{N}$ applications of the quantum oracle $U$ associated to the
function $f(x)$ defined in (\ref{csp}) are needed; we now show how
the problem can be solved within the sbit computational model.

\noindent {\it Sbit computational problem.} The sbit circuit,
corresponding to the classical circuit, implements the w--additive
operation $G: \K_ {n} \rightarrow \K_ {1}$ which maps the
computational basis state $\x$ of $\K_ {n}$ into $|f(x)\rrangle$,
where $f(x)$ is the function defined in (\ref{csp}). We now
describe the algorithm for solving the search problem.

{\it Step 1.} Initialize each sbit in the $\0$ state.

{\it Step 2.} Apply a H gate to the all sbits except to the last
one:
\begin{equation}
|000\ldots 00\rrangle \; \longrightarrow \; |sss \ldots s
0\rrangle \; = \; \sum_{x \in \A} \x,
\end{equation}
where $\A$ is the maximal set relative to $|sss \ldots
s0\rrangle$, i.e. the set of all $n$--bit--long strings whose last
bit is equal to $0$.

{\it Step 3.} Make a call to the circuit:
\begin{equation}
\sum_{x \in \A} \x \; \longrightarrow \; \sum_{x \in \A} G\, \x \;
= \; \left\{
\begin{array}{ll}
\s & \makebox{if $a \in \A$,} \\
\0 & \makebox{otherwise}
\end{array} \right.
\end{equation}

{\it Step 4.} Make a measurement; if the output is $\s$, then the
last bit of the binary expression of $a$ is equal to $0$;
otherwise it is equal to $1$: with just one call to the circuit we
have been able to find out the last digit of $a$.

{\it Step 5.} Repeat steps 2, 3 and 4 making the following change:
in step 2, apply an H gate to each input sbit except to the last
but one (the last but two, ...) A measurement of the output of the
circuit will reveal the value of the last but one (last but two,
...) bit of the binary expansion of $a$.

The above algorithm reaches the solution to the search problem
with just $n$ computations of the circuit, much faster that
Grover's algorithm.

As in the previous example, the {\it resources} needed to
implement the algorithm are {\it polynomial} in the size of the
problem: only $n$ sbits, $n(n-1)$ elementary gates and $n$
measurements are necessary; accordingly, our procedure is not
subject to the criticisms \cite{mey} raised against recent
proposals\footnote{See also refs. \cite{ahn,spr}, for experimental
realizations.} \cite{llo,gs} aiming at implementing Grover's
algorithm by resorting to classical mechanical systems.

\section{Implementation of the w--additive circuits}

In this section we face the problem of constructing the
w--additive circuit necessary for implementing the algorithms
previously discussed.

\subsection{Deutsch--Jozsa problem}

For simplicity we consider the simplest situation, in which we
have only one input bit (this is the original Deutsch problem
\cite{deu}). In this case, there are two {\it constant} functions:
\begin{equation} \label{cos}
f_{1}(x) \; = \; x\wedge\neg x\; \equiv \; 0 \qquad f_{2}(x) =
x\vee\neg x \; \equiv \; 1,
\end{equation}
and two {\it balanced} functions:
\begin{equation} \label{bal}
f_{3}(x) \; = \; x \qquad f_{4}(x) = \neg x;
\end{equation}
the classical circuits implementing $f_{1}(x)$ and $f_{2}(x)$ are
depicted in Fig. \ref{c1}, and those implementing $f_{3}(x)$ and
$f_{4}(x)$ are shown in Fig. \ref{c2}.
\begin{figure}
\begin{center}
\begin{picture}(220,120)(0,0)
\put(0,0){\line(1,0){220}} \put(0,120){\line(1,0){220}}
\put(0,0){\line(0,1){120}} \put(220,0){\line(0,1){120}}
\thicklines \put(40,100){\line(1,0){100}}
\put(80,80){\line(1,0){20}} \put(100,70){\line(0,1){20}}
\put(120,80){\line(-2,1){20}} \put(120,80){\line(-2,-1){20}}
\put(120,80){\line(1,0){20}} \put(140,75){\line(0,1){30}}
\qbezier(140,75)(190,90)(140,105) \put(165,90){\line(1,0){20}}
\put(80,100){\line(0,-1){20}} \put(80,100){\circle*{5}}
\put(40,40){\line(1,0){100}} \put(80,20){\line(1,0){20}}
\put(100,10){\line(0,1){20}} \put(120,20){\line(-2,1){20}}
\put(120,20){\line(-2,-1){20}} \put(120,20){\line(1,0){20}}
\qbezier(135,15)(150,30)(135,45) \qbezier(135,15)(190,30)(135,45)
\put(163,30){\line(1,0){20}} \put(80,40){\line(0,-1){20}}
\put(80,40){\circle*{5}}
\put(10,55){Circuit for $f_{1}(x)$} \put(140,55){Circuit for
$f_{2}(x)$}
\thinlines \qbezier(20,65)(20,80)(55,85)
\put(55,85){\vector(4,1){1}} \qbezier(200,45)(200,20)(180,20)
\put(180,20){\vector(-1,0){1}}
\end{picture}
\caption{Classical circuits for the two constant functions defined
in (\ref{cos}).} \label{c1}
\end{center}
\end{figure}
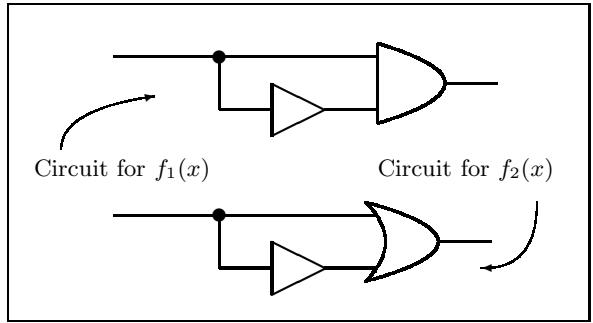
\begin{figure}
\begin{center}
\begin{picture}(220,120)(0,0)
\put(0,0){\line(1,0){220}} \put(0,120){\line(1,0){220}}
\put(0,0){\line(0,1){120}} \put(220,0){\line(0,1){120}}
\thicklines \put(40,90){\line(1,0){140}}
\put(40,30){\line(1,0){60}} \put(100,20){\line(0,1){20}}
\put(120,30){\line(-2,1){20}} \put(120,30){\line(-2,-1){20}}
\put(120,30){\line(1,0){60}}
\put(10,55){Circuit for $f_{3}(x)$} \put(140,55){Circuit for
$f_{4}(x)$}
\thinlines \qbezier(20,65)(20,75)(55,80)
\put(55,80){\vector(4,1){1}} \qbezier(200,45)(200,20)(180,20)
\put(180,20){\vector(-1,0){1}}
\end{picture}
\caption{Classical circuits for the two balanced functions defined
in (\ref{bal}).} \label{c2}
\end{center}
\end{figure}
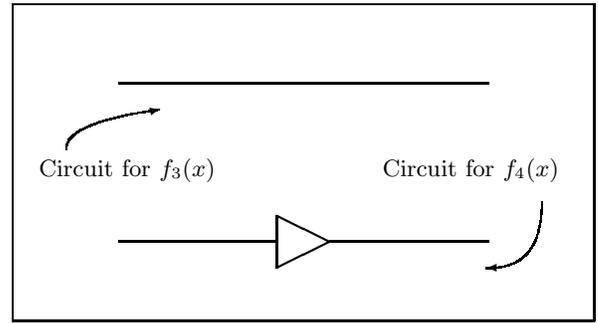

In all four cases, the classical circuits are convertible, i.e.
there is a procedure --- requiring a number of steps {\it
proportional} to the number of elementary gates --- for
constructing the corresponding w--additive circuits, which
consists in replacing every piece of the classical circuits by the
corresponding w--additive one. The w--additive circuits so
obtained, for completeness, are shown in Figs. \ref{w1} and
\ref{w2}.
\begin{figure}
\begin{center}
\begin{picture}(220,120)(0,0)
\put(0,0){\line(1,0){220}} \put(0,120){\line(1,0){220}}
\put(0,0){\line(0,1){120}} \put(220,0){\line(0,1){120}}
\thicklines \put(40,90){\line(1,0){55}}
\put(95,75){\line(0,1){30}} \put(125,75){\line(0,1){30}}
\put(95,75){\line(1,0){30}} \put(95,105){\line(1,0){30}}
\put(125,90){\line(1,0){55}} \put(106,87){$C_{0}$}
\put(40,30){\line(1,0){55}} \put(95,15){\line(0,1){30}}
\put(125,15){\line(0,1){30}} \put(95,15){\line(1,0){30}}
\put(95,45){\line(1,0){30}} \put(125,30){\line(1,0){55}}
\put(106,27){$C_{1}$}
\put(7,55){W-add. circuit for $f_{1}(x)$} \put(118,55){W-add.
circuit for $f_{2}(x)$}
\thinlines \qbezier(20,65)(20,75)(55,80)
\put(55,80){\vector(4,1){1}} \qbezier(200,45)(200,20)(180,20)
\put(180,20){\vector(-1,0){1}}
\end{picture}
\caption{W--additive circuits corresponding to the two circuits
depicted in Fig. \ref{c1}.} \label{w1}
\end{center}
\end{figure}
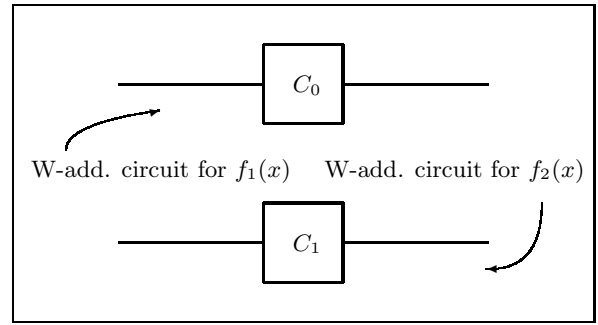
\begin{figure}
\begin{center}
\begin{picture}(220,120)(0,0)
\put(0,0){\line(1,0){220}} \put(0,120){\line(1,0){220}}
\put(0,0){\line(0,1){120}} \put(220,0){\line(0,1){120}}
\thicklines \put(40,90){\line(1,0){140}}
\put(40,30){\line(1,0){55}} \put(95,15){\line(0,1){30}}
\put(125,15){\line(0,1){30}} \put(95,15){\line(1,0){30}}
\put(95,45){\line(1,0){30}} \put(125,30){\line(1,0){55}}
\put(101,27){NOT}
\put(7,55){W-add. circuit for $f_{3}(x)$} \put(118,55){W-add.
circuit for $f_{4}(x)$}
\thinlines \qbezier(20,65)(20,75)(55,80)
\put(55,80){\vector(4,1){1}} \qbezier(200,45)(200,20)(180,20)
\put(180,20){\vector(-1,0){1}}
\end{picture}
\caption{W--additive circuits corresponding to the two circuits
depicted in Fig. \ref{c2}.} \label{w2}
\end{center}
\end{figure}
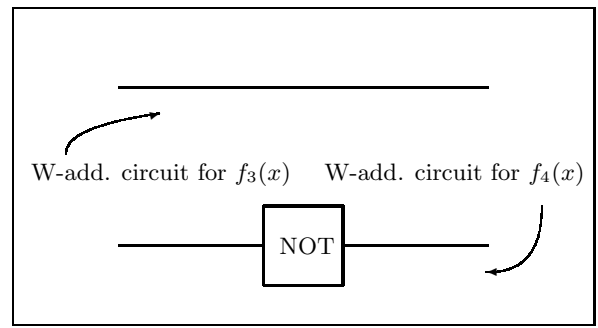

The crucial thing to analyze is how the w--additive circuits {\it
scale} when the domain of the constant/balanced functions
increases. Let us start with the constant functions. The two
constant functions taking $n$ bits into 1 bit can be written as
follows:
\begin{eqnarray}
f(x_{1},x_{2},\ldots x_{n}) & = & f_{1}(x_{1})\wedge
f_{1}(x_{2})\wedge \ldots f_{1}(x_{n}) \nonumber \\
g(x_{1},x_{2},\ldots x_{n}) & = & f_{2}(x_{1})\wedge
f_{2}(x_{2})\wedge \ldots f_{2}(x_{n}).
\end{eqnarray}
The unction $f$ is identically equal to 0, while $g$ is equal to
1; $f_{1}$ and $f_{2}$ are the functions defined in (\ref{cos}).

It is easy to see that $f$ and $g$ are implemented by circuits
which are convertible. Such circuits contain a number of
elementary gates which is proportional to $n$, the number of input
bits. This means that the procedure necessary to convert the
classical circuits into w--additive circuits requires a number of
steps proportional to $n$, and produces w--additive circuits whose
size is, again, proportional to $n$. Moreover, by construction,
such circuit act on the computational basis like their classical
counterparts.

In the case of balanced functions, the situation is more complex
since the number of possible functions increases with $n$. One way
to study the scaling problem is to recognize that the function
$f_{3}(x)$ previously defined is a particular example of a
``projection'' function. A projection $\pi_{j}$ is a function from
$n$ bits into 1 bit, which gives as the output the value of the
$j$--th bit\footnote{In our case, $f_{3}(x) \equiv \pi_{1}(x)$.};
e.g. $\pi_{3}(x_{1},x_{2},x_{3},x_{4}) = x_{3}$. In an analogous
way, the function $f_{4}(x)$ can be seen as the ``NOT'' of a
projection function. An example of a projection defined in terms
of Boolean connectives is the following:
\begin{eqnarray} \label{baf}
\lefteqn{\pi_{1}(x_{1},x_{2},\ldots x_{n}) \; =} \\
& & x_{1}\vee[(x_{2}\wedge\neg x_{2})\wedge (x_{3}\wedge\neg
x_{3})\wedge \ldots (x_{n}\wedge\neg x_{n})]. \nonumber
\end{eqnarray}
The function $\pi_{1}$ gives as the output the first bit $x_{1}$
and, clearly, it is balanced.

As in the case of constant functions, the classical circuit
implementing a balanced function of the type (\ref{baf}) is
convertible and its size is proportional to $n$: this means that,
once more, the procedure leading to the corresponding w--additive
circuit is also proportional to $n$ and produces a circuit whose
size is $n$.

This analysis shows that both the procedure leading from the
classical circuits to the corresponding w--additive circuits and
the size of the w--additive circuits scale {\it linearly} with
$n$, the number of input bits.

Of course, we stress once more that the great limitation of the S
model of computation is that we do not know whether all circuits
implementing balanced functions are convertible; anyway the
previous analysis shows that, within a precise mathematical
framework, some important features of quantum computation can be
recovered classically.

\subsection{Grover's search problem}

Grover's algorithm has found two important applications: it can
speed up the research through an unsorted database and it can
speedup the solution of NP--complete problems.  Let us start by
considering the search through an unsorted database. \\

\noindent {\bf Search through an unsorted database.} Suppose we
have an unsorted database (e.g. a list of names) containing $N$
elements, one of which is marked: we have to find that element. In
the literature, this search problem has been modelled as follows:
consider a quantum circuit implementing the unitary operator $U$
which, on the computational basis of ${\mathcal H}_{n}\otimes
{\mathcal H}_{1}$, maps $|x\rangle \otimes |y\rangle$ into
$|x\rangle \otimes |y\oplus f(x)\rangle$, where the function
$f(x)$ has been defined in equation (\ref{csp}). Once given the
circuit, Grovers' procedure allows one to find item $a$
quadratically faster than any classical algorithm. Needless to
say, no exponential growth is hidden into the construction of the
quantum circuit.

In ref. \cite{gs} a classical model mimicking Grover's algorithm
has been proposed. The authors consider a mechanical system of
coupled harmonic oscillators, all of which (but one) are equal:
each harmonic oscillator corresponds to an element of a database,
and the oscillator which is different from the others corresponds
to the item to be found. The authors show that one can resort to
Grover's algorithm for finding the desired oscillator, thus
obtaining a quadratic speedup over usual classical algorithms.

Is then possible to reproduce classically Grover's quantum search?
In ref. \cite{mey} it has been argued that this is not the case.
In fact, if the database contains $N$ elements (thus its size,
expressed in bits, is equal to $n = \log_{2}N$), the corresponding
mechanical system contains $2^n$ oscillators. As a consequence,
even if the search {\it algorithm} is as fast as the quantum one,
the overall procedure cannot be considered satisfactory from the
point of view of complexity theory, since the construction of the
{\it circuit} requires an exponential grow of physical resources.

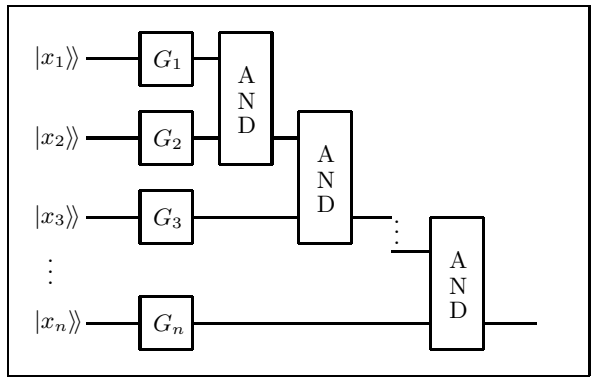
\begin{figure}
\begin{center}
\begin{picture}(220,140)(0,0)
\put(0,0){\line(1,0){220}} \put(0,140){\line(1,0){220}}
\put(0,0){\line(0,1){140}} \put(220,0){\line(0,1){140}}
\thicklines \put(30,120){\line(1,0){20}}
\put(30,90){\line(1,0){20}} \put(30,60){\line(1,0){20}}
\put(30,20){\line(1,0){20}}
\put(50,130){\line(1,0){20}} \put(50,110){\line(1,0){20}}
\put(50,110){\line(0,1){20}} \put(70,110){\line(0,1){20}}
\put(50,100){\line(1,0){20}} \put(50,80){\line(1,0){20}}
\put(50,80){\line(0,1){20}} \put(70,80){\line(0,1){20}}
\put(50,70){\line(1,0){20}} \put(50,50){\line(1,0){20}}
\put(50,50){\line(0,1){20}} \put(70,50){\line(0,1){20}}
\put(50,30){\line(1,0){20}} \put(50,10){\line(1,0){20}}
\put(50,10){\line(0,1){20}} \put(70,10){\line(0,1){20}}
\put(70,120){\line(1,0){10}} \put(70,90){\line(1,0){10}}
\put(70,60){\line(1,0){40}} \put(70,20){\line(1,0){90}}
\put(80,130){\line(1,0){20}} \put(80,80){\line(1,0){20}}
\put(80,80){\line(0,1){50}} \put(100,80){\line(0,1){50}}
\put(100,90){\line(1,0){10}}
\put(110,100){\line(1,0){20}} \put(110,50){\line(1,0){20}}
\put(110,50){\line(0,1){50}} \put(130,50){\line(0,1){50}}
\put(130,60){\line(1,0){15}}
\put(145,47){\line(1,0){15}}\put(160,60){\line(1,0){20}}
\put(160,10){\line(1,0){20}} \put(160,10){\line(0,1){50}}
\put(180,10){\line(0,1){50}} \put(180,20){\line(1,0){20}}
\put(10,118){$|x_{1}\rrangle$} \put(10,88){$|x_{2}\rrangle$}
\put(10,58){$|x_{3}\rrangle$} \put(10,18){$|x_{n}\rrangle$}
\put(55,117){$G_{1}$} \put(55,87){$G_{2}$} \put(55,57){$G_{3}$}
\put(55,17){$G_{n}$} \put(87,111){A}  \put(87,101){N}
\put(87,92){D} \put(117,81){A}  \put(117,71){N} \put(117,62){D}
\put(167,41){A}  \put(167,31){N} \put(167,22){D}
\put(15,35){$\vdots$} \put(146,50){$\vdots$}
\end{picture}
\caption{Circuit implementing the w--additive oracle for Grover's
search problem; the theorems of the previous section ensure that
the circuit is w--additive. The gate $G_{i}$ applied to sbit $i$
is the identity gate if $a_{i} = 1$, otherwise it is a NOT gate:
there are precisely $2^n$ ways to arrange identity and NOT gates
in the circuit, each configuration corresponding to a different
oracle, i.e. to a different function of the type (\ref{csp}).}
\label{fig2}
\end{center}
\end{figure}

By resorting to the S model of computation, we have proposed an
algorithm which is even faster than Grover's. What about the size
of the w--additive circuit? Here we show that no exponential
growth of physical resources is hidden in it. An example of
w--additive circuit taking $n$ sbits into 1 sbit, which maps the
state $|x\rrangle$ of the computational basis of ${\mathcal
K}_{n}$ into $|f(x)\rrangle$, is shown in Fig. \ref{fig2}.
The circuit contains $2n-1$ elementary gates, thus its size grows
{\it linearly} with the number of bits encoding the size of the
problem. Moreover, the circuit does not contain loops or any other
trick hiding an exponential growth, e.g., in the time it requires
to perform a computation: the number of physical resources is
genuinely proportional to $n$.

But then we arrive to a paradox: how can the S model of
computation, which is essentially classical, do better than
classical computation? The answer is simple: in order to construct
the w--additive circuit, we have to {\it load} the classical
database (i.e. the list of names) into the hypothetical
w--additive computer, and this operation requires a number of
steps equal to $N = 2^n$. Thus, there is no contradiction with
well known classical results. Note anyway that the loading
procedure is necessary also within quantum computation, in order
to construct the quantum counterpart of the classical database.

Accordingly, if one does not take into account the loading
procedure, then the S model of computation provides a fast way of
searching through a database, and no hidden exponential growth of
physical resources or time or energy is present. The only
exponential slowdown appears in the loading procedure: but this is
common to other situations, like quantum computation. \\

\noindent {\bf NP--complete problems.} Let us consider again the
SAT problem: given a classical Boolean circuit taking $n$ bits
into one bit, we have do determined whether or not there is an
input value whose output is 1. As already pointed out, the
procedure for building the quantum analog of a classical circuit
is polynomial in the size of the problem (i.e. in the number $n$
of elementary gates forming the classical circuit); once one has
the quantum circuit, he can use (a slight modification of)
Grover's algorithm for finding the solution of the problem.

Within the S model of computation the situation is different. As
we have seen in the previous section, once given the appropriate
w--additive circuit, there is a fast algorithm for finding a
solution of the SAT problem. Anyway, in general we do {\it not}
known of a {\it general efficient} procedure for transforming a
classical circuit into the corresponding w--additive one, to which
the algorithm can be applied. Only when the classical circuit is
{\it convertible}, like the one depicted in Fig. \ref{ex}, we can
easily construct the corresponding w--additive circuit and apply
the search algorithm to it, and the overall procedure is
efficient. But, as already said, it is an open problem to
determine which classical circuits are convertible.

\section{Physical implementation of the sbit model of computation}

Sbit--computation can be physically implemented in various ways:
here we propose a simple one. The state of a {\it single} sbit is
associated to 2 bits, according to the rules:
\begin{eqnarray}
\0 & \longrightarrow & 10 \nonumber \\
\1 & \longrightarrow & 01 \nonumber \\
\s & \longrightarrow & 11 \nonumber;
\end{eqnarray}
the state 00 is not taken into account. Thus, physically, a {\it
sbit} is realized in terms of two wires with current passing
through them. In a similar way, the state of $n$ sbits is
associated to $n$ couples of classical bits, and is physically
realized by $n$ couples of wires, according to the previous rules.

{\it W--additive gates} are easily implemented in terms of
classical gates: as an example, Figs. \ref{not}, \ref{fand} and
\ref{for} show the classical circuits for the w--additive NOT, OR,
and AND gates.
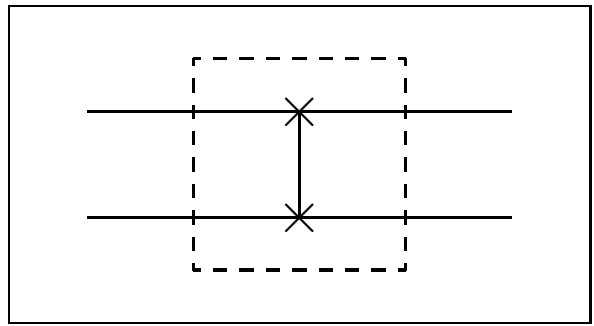
\begin{figure}
\begin{center}
\begin{picture}(220,120)(0,0)
\put(0,0){\line(1,0){220}} \put(0,120){\line(1,0){220}}
\put(0,0){\line(0,1){120}} \put(220,0){\line(0,1){120}}
\thicklines \put(30,40){\line(1,0){160}}
\put(30,80){\line(1,0){160}} \put(110,40){\line(0,1){40}}
\put(105,35){\line(1,1){10}} \put(115,35){\line(-1,1){10}}
\put(105,75){\line(1,1){10}} \put(115,75){\line(-1,1){10}}
\put(70,20){\dashbox{5}(80,80){ }}
\end{picture}
\caption{Circuit for the w--additive NOT gate: it is a classical
swap gate, which interchanges the values of the two input bits.}
\label{not}
\end{center}
\end{figure}
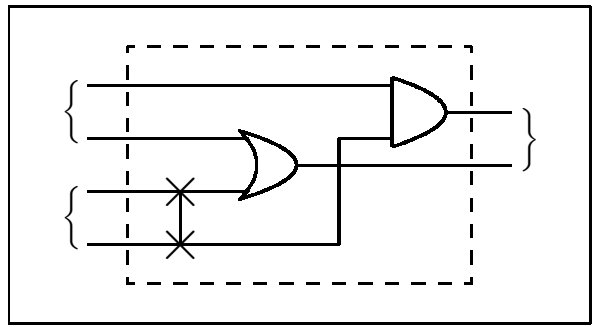
\begin{figure}
\begin{center}
\begin{picture}(220,120)(0,0)
\put(0,0){\line(1,0){220}} \put(0,120){\line(1,0){220}}
\put(0,0){\line(0,1){120}} \put(220,0){\line(0,1){120}}
\thicklines \put(30,30){\line(1,0){95}}
\put(30,50){\line(1,0){60}} \put(30,70){\line(1,0){60}}
\put(30,90){\line(1,0){115}}
\put(65,30){\line(0,1){20}} \put(60,25){\line(1,1){10}}
\put(70,25){\line(-1,1){10}} \put(60,45){\line(1,1){10}}
\put(70,45){\line(-1,1){10}}
\qbezier(87,47)(100,60)(87,73) \qbezier(87,47)(130,60)(87,73)
\put(109,60){\line(1,0){81}}
\put(125,30){\line(0,1){40}} \put(125,70){\line(1,0){20}}
\put(145,67){\line(0,1){26}} \qbezier(145,67)(185,80)(145,93)
\put(165,80){\line(1,0){25}}
\put(45,15){\dashbox{5}(130,90){ }} \put(20,78){$\left\{
\makebox(8,15)[t]{$ $} \right.$} \put(20,38){$\left\{
\makebox(8,15)[t]{$ $} \right.$} \put(184,67.5){$\left.
\makebox(8,15)[t]{$ $} \right\}$}
\end{picture}
\caption{Circuit for the w--additive OR gate: a classical swap
gate is applied to the last two bits, followed by a classical OR
gate between the second and the third bit and a classical AND gate
between the first and fourth bit.} \label{fand}
\end{center}
\end{figure}
\begin{figure}
\begin{center}
\begin{picture}(220,120)(0,0)
\put(0,0){\line(1,0){220}} \put(0,120){\line(1,0){220}}
\put(0,0){\line(0,1){120}} \put(220,0){\line(0,1){120}}
\thicklines \put(30,30){\line(1,0){95}}
\put(30,50){\line(1,0){60}} \put(30,70){\line(1,0){60}}
\put(30,90){\line(1,0){115}}
\put(65,30){\line(0,1){20}} \put(60,25){\line(1,1){10}}
\put(70,25){\line(-1,1){10}} \put(60,45){\line(1,1){10}}
\put(70,45){\line(-1,1){10}}
\qbezier(90,47)(130,60)(90,73) \put(90,47){\line(0,1){26}}
\put(110,60){\line(1,0){80}}
\put(125,30){\line(0,1){40}} \put(125,70){\line(1,0){20}}
\qbezier(142,67)(155,80)(142,93) \qbezier(142,67)(185,80)(142,93)
\put(164,80){\line(1,0){26}}
\put(45,15){\dashbox{5}(130,90){ }} \put(20,78){$\left\{
\makebox(8,15)[t]{$ $} \right.$} \put(20,38){$\left\{
\makebox(8,15)[t]{$ $} \right.$} \put(184,67.5){$\left.
\makebox(8,15)[t]{$ $} \right\}$}
\end{picture}
\caption{Circuit for the w--additive AND gate: a classical swap
gate is applied to the last two bits, followed by a classical AND
gate between the second and the third bit and a classical OR gate
between the first and fourth bit.} \label{for}
\end{center}
\end{figure}
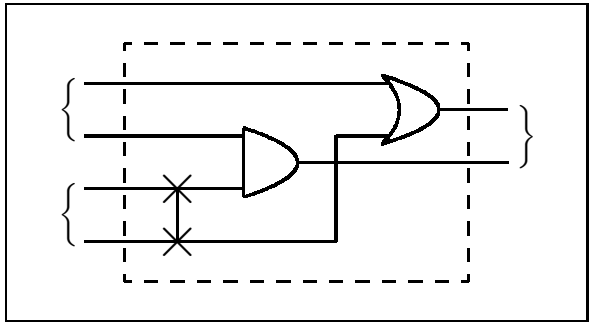

According to the previous rules, any w--additive circuit with $n$
input sbits and composed of $m$ elementary w--additive gates can
be implemented by a classical circuit working on $2n$ bits and
made up of $k\cdot m$ elementary classical gates, $k$ being a
fixed constant: this means that the sbit model of computation is
polynomially reducible to the classical circuit model of
computation, and thus {\it equivalent} to it.

\section{Discussion and perspectives}

In the previous sections we have developed the S model of
computation: this is a well defined computational model which
presents some of the characteristic traits of quantum computation;
in particular, there are states which are {\it superpositions} (to
be understood in the appropriate way) of the computational basis
states. Among the possible gates, the only allowed ones are those
which satisfy the {\it w--additivity} condition. Thanks to this
property (which is the analog of quantum linearity), one needs to
define and control operators only on the computational basis of
their domain, and w--additivity automatically defines their action
on all other states.

We have divided a computational problem into two parts: the
construction of the circuit which implements the function $f(x)$
defining a problem, and the identification of the algorithm for
finding the solution of the problem. We have shown that the S
model allows the working out of algorithms which are faster than
the corresponding classical and quantum algorithms; on the other
hand, no general procedure for transforming a classical circuit
into the corresponding w--additive circuit is known (while any
classical circuit can be efficiently transformed into a quantum
circuit): only convertible circuit admit --- by definition ---
easy procedures for constructing the corresponding w--additive
circuits.

From the previous analysis we can make the following comments on
quantum computation:
\begin{enumerate}
\item The original Deutsch problem \cite{deu} of distinguishing
between the two constant function (\ref{cos}) and the two balanced
functions (\ref{bal}) can be solved also classically with a single
query to the appropriate w--additive circuit. Moreover, the global
procedure scales polynomially with the size of the problem.

\item Similarly, the search problem of an unsorted database can be
solved efficiently by the S (classical) model of computation,
modulo the loading procedure of the database.

\item Since the problems one is interested in solving are, in
general, classical problems defined in terms of classical
functions, any discussion on the efficiency of a quantum algorithm
must take into account also the procedure for constructing the
quantum circuit implementing the classical function. In
particular, all discussions about the power of quantum computation
for problems involving oracles should be made more precise by
analyzing in detail the procedure necessary for converting a
classical oracle into a quantum one.
\end{enumerate}
This last point is of particular relevance. In the literature,
classical algorithms involving an oracle are compared with quantum
algorithms involving the corresponding quantum oracle. In the
light of the previous analysis, one could also consider
w--additive algorithms which resort to w--additive oracles: there
would be no surprise to find out that, relative to an oracle, many
other w--additive algorithm exist which are faster than the
corresponding classical and quantum ones. This means that, in
order to make the discussion clear and rigorous, one must always
take into account the procedure needed to construct the quantum
(or w--additive) oracles out of the classical one.

\section*{Acknowledgements}

We are indebted to S. Azaele, A. Braunstein, D. D\"urr, S.
Goldstein, L. Marinatto, P. Solinas, Nino Zangh\`{\i} and R.
Zecchina for very stimulating and useful discussions.

\section{Appendix 1: proof of theorems}

\noindent {\bf Theorem:} there exists a universal set of
elementary w--additive gates.

\noindent {\it Proof.} We have to show that, by resorting to a
{\it fixed} number of w--additive gates, we can compute {\it any}
w--additive operation
\begin{equation}
G: \K_ {n} \; \longrightarrow \; \K_ {m}.
\end{equation}
Since a operator with a $m$--sbit value is equivalent to $m$
operators with one--sbit value each, it suffices to show that a
fixed number of gates is sufficient to compute all w--additive
operations
\begin{equation}
G: \K_ {n} \; \longrightarrow \; \K_ {1}.
\end{equation}
The proof is by induction on the ``dimension'' $n$ of the domain.
For $n = 1$, there are 9 w--additive gates: the identity gate
(which does nothing on the input), the NOT and H gates, and the
following w--additive gates:
\begin{center}
C$_{0}$ =
\begin{tabular}{c|c}
input & output \\\hline
$\0$ & $\0$ \\
$\1$ & $\0$
\end{tabular}
\hspace{1cm} C$_{1}$  =
\begin{tabular}{c|c}
input & output \\\hline
$\0$ & $\1$ \\
$\1$ & $\1$
\end{tabular}
\end{center}
\begin{center}
S$_{0}$ =
\begin{tabular}{c|c}
input & output \\\hline
$\0$ & $\s$ \\
$\1$ & $\0$
\end{tabular}
\hspace{1cm} $\overline{\makebox{S}}_{0}$  =
\begin{tabular}{c|c}
input & output \\\hline
$\0$ & $\0$ \\
$\1$ & $\s$
\end{tabular}
\end{center}
\begin{center}
S$_{1}$ =
\begin{tabular}{c|c}
input & output \\\hline
$\0$ & $\s$ \\
$\1$ & $\1$
\end{tabular}
\hspace{1cm} $\overline{\makebox{S}}_{1}$  =
\begin{tabular}{c|c}
input & output \\\hline
$\0$ & $\1$ \\
$\1$ & $\s$
\end{tabular}
\end{center}
All the above mentioned gates can be expressed in terms of NOT,
AND, OR and S$_{0}$ gates. As a matter of fact, C$_{0}$ can be
implemented using an AND gate with the second sbit set equal to
$\0$; C$_{1}$ is equivalent to a OR gate with the second sbit
equal to $\1$; $\overline{\makebox{S}}_{0}$ is equivalent to a NOT
gate followed by a S$_{0}$ gate; $\overline{\makebox{S}}_{1}$ is
equivalent to a NOT gate followed by a S$_{1}$ gate, while S$_{1}$
is equivalent to a S$_{0}$ gate followed by a NOT gate; finally, H
is equivalent to a S$_{1}$ gate followed by a
$\overline{\makebox{S}}_{1}$ gate.

Assume now that any w--additive operator on $n$ sbits can be
computed by a circuit consisting of w--additive elementary gates,
and consider the w--additive operator
\begin{equation}
G: \K_ {n+1} \; \longrightarrow \; \K_ {1},
\end{equation}
having $n+1$ sbits as input. Let us introduce the following two
operations taking $n$ sbits into one sbit:
\begin{eqnarray}
G_{0} |a_{1} a_{2} \ldots a_{n} \rrangle & = & G |0 a_{1} a_{2}
\ldots a_{n} \rrangle \label{gzero} \\
G_{1} |a_{1} a_{2} \ldots a_{n} \rrangle & = & G |1 a_{1} a_{2}
\ldots a_{n} \rrangle. \label{guno}
\end{eqnarray}
It is easy to show that they are w--additive, so by the inductive
hypothesis there exists w--additive circuits computing them. The
following relation follows from (\ref{gzero}) and (\ref{guno}):
\begin{eqnarray} \label{gsum}
G |s a_{1} a_{2} \ldots a_{n} \rrangle & = & G_{0} |a_{1} a_{2}
\ldots a_{n} \rrangle + \nonumber \\
& + & G_{1} |a_{1} a_{2} \ldots a_{n} \rrangle,
\end{eqnarray}
for any state $|a_{1} a_{2} \ldots a_{n} \rrangle \in \K_ {n}$.

Let us now consider the w--additive T gate, taking three sbits
into one sbit, which is defined on the computational basis in the
following way:
\begin{center}
T \hspace{.2cm} = \hspace{.2cm}
\begin{tabular}{c|c}
input & output \\\hline
$|000\rrangle$ & $\0$ \\
$|001\rrangle$ & $\0$ \\
$|010\rrangle$ & $\1$ \\
$|011\rrangle$ & $\1$
\end{tabular} \hspace{.5cm}
\begin{tabular}{c|c}
input & output \\\hline
$|100\rrangle$ & $\0$ \\
$|101\rrangle$ & $\1$ \\
$|110\rrangle$ & $\0$ \\
$|111\rrangle$ & $\1$
\end{tabular} \hspace{.5cm}
\end{center}
T has one control sbit (the first) and two target sbits: if the
control sbit is set to $\0$, then the output is equal to the first
target sbit, i.e. the second sbit of the row; if on the other hand
the control sbit is equal to $\1$, the output is equal to the
second target sbit. It is easy to check that the w--additivity
condition preserves the above property, e.g. $|0s1\rrangle$ is
mapped into $\s$ and $|1s1\rrangle$ into $\1$; moreover, if the
control sbit is equal to $\s$, the output is equal to the sum of
the two target sbits.

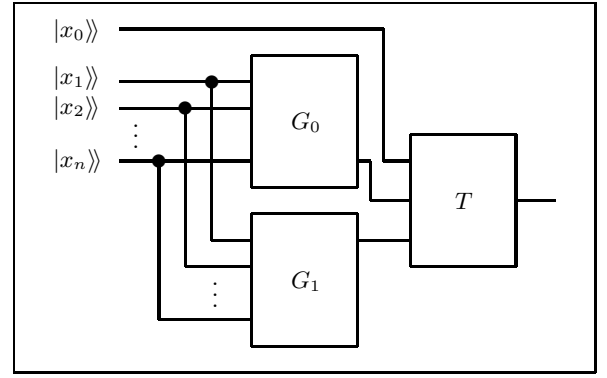
\begin{figure}
\begin{center}
\begin{picture}(220,130)(0,0)
\put(0,0){\line(1,0){220}} \put(0,140){\line(1,0){220}}
\put(0,0){\line(0,1){140}} \put(220,0){\line(0,1){140}}
\thicklines \put(40,130){\line(1,0){100}}
\put(140,130){\line(0,-1){50}} \put(40,110){\line(1,0){50}}
\put(75,110){\line(0,-1){60}} \put(75,110){\circle*{5}}
\put(40,100){\line(1,0){50}} \put(65,100){\line(0,-1){60}}
\put(65,100){\circle*{5}} \put(40,80){\line(1,0){50}}
\put(55,80){\line(0,-1){60}} \put(55,80){\circle*{5}}
\put(75,50){\line(1,0){15}} \put(65,40){\line(1,0){25}}
\put(55,20){\line(1,0){35}}
\put(90,70){\line(1,0){40}} \put(90,120){\line(1,0){40}}
\put(90,70){\line(0,1){50}} \put(130,70){\line(0,1){50}}
\put(90,10){\line(1,0){40}} \put(90,60){\line(1,0){40}}
\put(90,10){\line(0,1){50}} \put(130,10){\line(0,1){50}}
\put(140,80){\line(1,0){10}} \put(130,80){\line(1,0){5}}
\put(135,80){\line(0,-1){15}} \put(135,65){\line(1,0){15}}
\put(130,50){\line(1,0){20}} \put(150,40){\line(1,0){40}}
\put(150,90){\line(1,0){40}} \put(150,40){\line(0,1){50}}
\put(190,40){\line(0,1){50}} \put(190,65){\line(1,0){15}}
\put(15,127){$|x_{0}\rrangle$} \put(15,109){$|x_{1}\rrangle$}
\put(15,98){$|x_{2}\rrangle$}  \put(15,78){$|x_{n}\rrangle$}
\put(105,92){$G_{0}$} \put(105,32){$G_{1}$}  \put(167,62){$T$}
\put(45,85){$\vdots$} \put(75,25){$\vdots$}
\end{picture}
\caption{Circuit for computing an arbitrary w--additive operation
$G$ on $n+1$ sbits, assuming that there are w--additive circuits
for computing the $n$ sbits operations $G_{0}$ and $G_{1}$.}
\label{fig1}
\end{center}
\end{figure}

With the help of the T gate, it is easy to devise a circuit that
computes G (see Fig. \ref{fig1}). The circuit computes both
G$_{0}$ and G$_{1}$ on the last $n$ sbits; then, depending on
whether the first sbit is equal to $\0$, $\1$ or $\s$, the T gate
outputs G$_{0}$, G$_{1}$ or G$_{0}$ + G$_{1}$, which is the
desired outcome, according to Eqs. (\ref{gzero}), (\ref{guno}) and
(\ref{gsum}).

The above proof shows that we can identify the following set of
universal gates: NOT, S$_{0}$, FANOUT, AND, OR and T gate.

\noindent {\bf Theorem:} let $G_{1} : \K_ {n} \rightarrow \K_ {1}$
and $G_{2} : \K_ {m} \rightarrow \K_ {1}$ be two w--additive
operations. Then $G : \K_ {n} \times \K_ {m} \rightarrow \K_ {1}$
defined as $G \equiv$ AND $[ G_{1} \times G_{2} ]$ is also
w--additive.

\noindent {\it Proof.} Let us consider a generic $|a\rrangle \in
\K_ {n} \times \K_ {m}$,
\begin{equation}
|a\rrangle = \sum_{x \in {\mathcal A}} |x\rrangle.
\end{equation}
$|a\rrangle$ can also be written in the following way:
\begin{eqnarray}
|a\rrangle = |a_{1}\rrangle\,|a_{2}\rrangle, & \quad &
|a_{1}\rrangle = \sum_{x_{1} \in {\mathcal A}_{1}}
|x_{1}\rrangle, \\
& & |a_{2}\rrangle = \sum_{x_{2} \in {\mathcal A}_{2}}
|x_{2}\rrangle, \nonumber
\end{eqnarray}
with $|a_{1}\rrangle, |x_{1}\rrangle \in \K_ {n}$ and
$|a_{2}\rrangle, |x_{2}\rrangle \in \K_ {m}$. It is easy to prove
that the set ${\mathcal A}$ is the direct product of the two sets
${\mathcal A}_{1}$ and ${\mathcal A}_{2}$; accordingly, any
element $|x\rrangle \in {\mathcal A}$ can be written as
$|x\rrangle = |x_{1}\rrangle\, |x_{2}\rrangle$, with
$|x_{1}\rrangle \in {\mathcal A}_{1}$ and $|x_{2}\rrangle \in
{\mathcal A}_{2}$, and viceversa. We recall that ${\mathcal A}$,
${\mathcal A}_{1}$ and ${\mathcal A}_{2}$ are the {\it maximal}
sets associated to $|a\rrangle$, $|a_{1}\rrangle$ and
$|a_{2}\rrangle$, respectively.

It is convenient to distinguish the following three cases.

\noindent First case: $\forall\; x_{1} \in {\mathcal A}_{1}, \;
G_{1} |x_{1}\rrangle = \0$ or $\forall\; x_{2} \in {\mathcal
A}_{2}, \; G_{2} |x_{2}\rrangle = \0$; suppose that the first
situation is true. Resorting to the w-additivity of $G_{1}$, we
have:
\begin{eqnarray}
G |x\rrangle & = & \makebox{AND}\, [ G_{1} |x_{1}\rrangle \, G_{2}
|x_{2}\rrangle ] = \0 \quad \forall\; x \in {\mathcal A}, \nonumber \\
G |a\rrangle & = & \makebox{AND}\, [ G_{1} |a_{1}\rrangle \, G_{2}
|a_{2}\rrangle ] = \0;
\end{eqnarray}
this means the $G$ is w--additive.

\noindent Second case: $\forall\; x_{1} \in {\mathcal A}_{1}, \;
G_{1} |x_{1}\rrangle = \1$ and $\forall\; x_{2} \in {\mathcal
A}_{2}, \; G_{2} |x_{2}\rrangle = \1$; since both $G_{1}$ and
$G_{2}$ are w--additive operators, we get:
\begin{eqnarray}
G |x\rrangle & = & \makebox{AND}\, [ G_{1} |x_{1}\rrangle \, G_{2}
|x_{2}\rrangle ] = \1 \quad \forall\; x \in {\mathcal A}, \nonumber \\
G |a\rrangle & = & \makebox{AND}\, [ G_{1} |a_{1}\rrangle \, G_{2}
|a_{2}\rrangle ] = \1;
\end{eqnarray}
also in this case $G$ is w--additive.

\noindent Third case:  $(\exists\, \overline{x}_{1} \in {\mathcal
A}_{1}, \exists\, \overline{x}_{2} \in {\mathcal A}_{2}$ such that
$G_{1} |\overline{x}_{1}\rrangle \neq \0, G_{2}
|\overline{x}_{2}\rrangle \neq \0)$ and $(\exists\,
\overline{y}_{1} \in {\mathcal A}_{1}$ such that $G_{1}
|\overline{y}_{1}\rrangle \neq \1$ or $\exists\, \overline{y}_{2}
\in {\mathcal A}_{2}$ such that $G_{2} |\overline{x}_{2}\rrangle
\neq \1)$, i.e. both case 1 and case 2 are excluded. The above
conditions imply that both $G_{1} |a_{1} \rrangle \neq \0$ and
$G_{2} |a_{2} \rrangle \neq \0$, and $G_{1} |a_{1} \rrangle \neq
\1$ or $G_{2} |a_{2} \rrangle \neq \1$, which in turn implies that
$G |a\rrangle = |s\rrangle$.

Assume now that $G$ is not w--additive,
\begin{equation}
G |a\rrangle = |s\rrangle \neq \sum_{x \in {\mathcal A}} G
|x\rrangle,
\end{equation}
which means that either $G |x\rrangle = \1$ $\forall\; x \in
{\mathcal A}$ or $G |x\rrangle = \0$ $\forall\; x \in {\mathcal
A}$. Let us consider the case in which $G |x\rrangle = \1$
$\forall\; x \in {\mathcal A}$; then we have that $\forall\,
|x\rrangle \in {\mathcal A}, |x\rrangle = |x_{1}\rrangle
|x_{2}\rrangle$: AND $[G_{1} |x_{1}\rrangle G_{2} |x_{1}\rrangle]
= \1$, which implies that both $G_{1} |x_{1}\rrangle = 1 \;
\forall\, x_{1} \in {\mathcal A}_{1}$ and $G_{2} |x_{2}\rrangle =
1 \; \forall\, x_{2} \in {\mathcal A}_{2}$. But this cannot happen
since such a situation (corresponding to case 2) has been
excluded.

If on the other hand $G |x\rrangle = \0$ $\forall\; x \in
{\mathcal A}$, we have that for any couple $x_{1} \in {\mathcal
A}_{1}$, $x_{2} \in {\mathcal A}_{2}$: $G_{1} |x_{1}\rrangle = \0$
or $G_{2} |x_{2}\rrangle = \0$. This necessarily implies that
$G_{1}$ is constant and equal to $\0$ or that $G_{2}$ is constant
and equal to $\0$ (in fact, if this were not true we would have
$\exists\, \overline{x}_{1} \in {\mathcal A}_{1}, \exists\,
\overline{x}_{2} \in {\mathcal A}_{2}$ such that both $G_{1}
|\overline{x}_{1}\rrangle \neq \0$ and $G_{2}
|\overline{x}_{2}\rrangle \neq \0$ which negates the previous
statement) but, again, this situation (corresponding to case 1)
has been excluded. The conclusion is that also for case 3 the
operator $G$ must be w--additive. This completes the proof.

\section{Appendix 2: gates}

We list all the w--additive gates which have been introduced in
the paper, writing explicitly their action on the entire domain.

One--sbit gates:
\begin{center}
I =
\begin{tabular}{c|c}
input & output \\\hline
$\0$ & $\0$ \\
$\1$ & $\1$ \\\hline $\s$ & $\s$
\end{tabular}
\hspace{1cm} NOT =
\begin{tabular}{c|c}
input & output \\\hline
$\0$ & $\1$ \\
$\1$ & $\0$ \\\hline $\s$ & $\s$
\end{tabular}
\end{center}
\begin{center}
C$_{0}$ =
\begin{tabular}{c|c}
input & output \\\hline
$\0$ & $\0$ \\
$\1$ & $\0$ \\\hline $\s$ & $\0$
\end{tabular}
\hspace{1cm} C$_{1}$  =
\begin{tabular}{c|c}
input & output \\\hline
$\0$ & $\1$ \\
$\1$ & $\1$ \\\hline $\s$ & $\1$
\end{tabular}
\end{center}
\begin{center}
S$_{0}$ =
\begin{tabular}{c|c}
input & output \\\hline
$\0$ & $\s$ \\
$\1$ & $\0$ \\\hline $\s$ & $\s$
\end{tabular}
\hspace{1cm} $\overline{\makebox{S}}_{0}$  =
\begin{tabular}{c|c}
input & output \\\hline
$\0$ & $\0$ \\
$\1$ & $\s$ \\\hline $\s$ & $\s$
\end{tabular}
\end{center}
\begin{center}
S$_{1}$ =
\begin{tabular}{c|c}
input & output \\\hline
$\0$ & $\s$ \\
$\1$ & $\1$ \\\hline $\s$ & $\s$
\end{tabular}
\hspace{1cm} $\overline{\makebox{S}}_{1}$  =
\begin{tabular}{c|c}
input & output \\\hline
$\0$ & $\1$ \\
$\1$ & $\s$ \\\hline $\s$ & $\s$
\end{tabular}
\end{center}
\begin{center}
H =
\begin{tabular}{c|c}
input & output \\\hline
$\0$ & $\s$ \\
$\1$ & $\s$ \\\hline $\s$ & $\s$
\end{tabular}
\hspace{.1cm} FANOUT =
\begin{tabular}{c|c}
input & output \\\hline
$\0$ & $|00\rrangle$ \\
$\1$ & $|11\rrangle$ \\\hline $\s$ &
$|\makebox{s}\makebox{s}\rrangle$
\end{tabular}
\end{center}
Two--sbit gates:
\begin{center}
\begin{tabular}{ccc}
AND \hspace{.7cm} & OR \hspace{.7cm} & XOR \hspace{.1cm} \\
\begin{tabular}{c|c}
input & output \\\hline
$|00\rrangle$ & $\0$ \\
$|01\rrangle$ & $\0$ \\
$|10\rrangle$ & $\0$ \\
$|11\rrangle$ & $\1$ \\\hline $|0\makebox{s}\rrangle$ & $\0$ \\
$|1s\rrangle$ & $\s$ \\
$|s0\rrangle$ & $\0$ \\
$|s1\rrangle$ & $\s$ \\
$|ss\rrangle$ & $\s$
\end{tabular} \hspace{.5cm}
&
\begin{tabular}{c|c}
input & output \\\hline
$|00\rrangle$ & $\0$ \\
$|01\rrangle$ & $\1$ \\
$|10\rrangle$ & $\1$ \\
$|11\rrangle$ & $\1$ \\\hline $|0\makebox{s}\rrangle$ & $\s$ \\
$|1s\rrangle$ & $\1$ \\
$|s0\rrangle$ & $\s$ \\
$|s1\rrangle$ & $\1$ \\
$|ss\rrangle$ & $\s$
\end{tabular} \hspace{.5cm}
&
\begin{tabular}{c|c}
input & output \\\hline
$|00\rrangle$ & $\0$ \\
$|01\rrangle$ & $\1$ \\
$|10\rrangle$ & $\1$ \\
$|11\rrangle$ & $\0$ \\\hline $|0\makebox{s}\rrangle$ & $\s$ \\
$|1s\rrangle$ & $\s$ \\
$|s0\rrangle$ & $\s$ \\
$|s1\rrangle$ & $\s$ \\
$|ss\rrangle$ & $\s$
\end{tabular}
\end{tabular}
\end{center}
The T gate:
\begin{center}
\begin{tabular}{ccc}
\begin{tabular}{c|c}
input & output \\\hline
$|000\rrangle$ & $\0$ \\
$|001\rrangle$ & $\0$ \\
$|00s\rrangle$ & $\0$ \\
$|010\rrangle$ & $\1$ \\
$|011\rrangle$ & $\1$ \\
$|01s\rrangle$ & $\1$ \\
$|0s0\rrangle$ & $\s$ \\
$|0s1\rrangle$ & $\s$ \\
$|0ss\rrangle$ & $\s$
\end{tabular} \hspace{.5cm}
&
\begin{tabular}{c|c}
input & output \\\hline
$|100\rrangle$ & $\0$ \\
$|101\rrangle$ & $\1$ \\
$|10s\rrangle$ & $\s$ \\
$|110\rrangle$ & $\0$ \\
$|111\rrangle$ & $\1$ \\
$|11s\rrangle$ & $\s$ \\
$|1s0\rrangle$ & $\0$ \\
$|1s1\rrangle$ & $\1$ \\
$|1ss\rrangle$ & $\s$
\end{tabular} \hspace{.5cm}
&
\begin{tabular}{c|c}
input & output \\\hline
$|s00\rrangle$ & $\0$ \\
$|s01\rrangle$ & $\s$ \\
$|s0s\rrangle$ & $\s$ \\
$|s10\rrangle$ & $\s$ \\
$|s11\rrangle$ & $\1$ \\
$|s1s\rrangle$ & $\s$ \\
$|ss0\rrangle$ & $\s$ \\
$|ss1\rrangle$ & $\s$ \\
$|sss\rrangle$ & $\s$
\end{tabular}
\end{tabular}
\end{center}

\end{document}